\documentclass{IEEEtran}
\IEEEoverridecommandlockouts
\usepackage{cite}
\usepackage{amsmath,amssymb,amsfonts}
\usepackage{algorithmic}
\usepackage{graphicx}
\usepackage{textcomp}
\usepackage{xcolor}
\usepackage{setspace}
\usepackage{enumitem}
\def\BibTeX{{\rm B\kern-.05em{\sc i\kern-.025em b}\kern-.08em
		T\kern-.1667em\lower.7ex\hbox{E}\kern-.125emX}}
\usepackage{subfigure}
\newtheorem{my_theorem}{Theorem}

\title{Direct Air-to-Underwater Optical Wireless Communication: Statistical Characterization and Outage Performance }
\author{Ziyaur Rahman, ~\IEEEmembership{Graduate Student Member,~IEEE}, S.~M.~ Zafaruddin,~\IEEEmembership{Senior Member,~IEEE}, and
	V.~K.~ Chaubey,~\IEEEmembership{Senior Member,~IEEE} 
	\thanks{This work was supported in part by the 
		 Science and Engineering Research Board (SERB), India under MATRICS Grant MTR/2021/000890 and Start-up Research Grant SRG/2019/002345.
		
The authors	are  with  the Department of Electrical and Electronics Engineering, Birla Institute of Technology and Science, Pilani, Pilani-333031, Rajasthan, India. (Email: \{p20170416, syed.zafaruddin, vkc\}@pilani.bits-pilani.ac.in.}
	
	\thanks{}}

\begin{document}
	\maketitle

\begin{abstract}\label{sec:abstract}
In general, a  buoy relay  is used to connect the underwater communication to the terrestrial network over a radio or optical wireless communication (OWC) link. The use of relay deployment  may pose security and deployment issues.  This paper investigates the feasibility of   direct air-to-underwater (A2UW) communication  from an over-the-sea OWC system  to an underwater submarine without deploying a relaying node. We analyze the statistical performance of the direct transmission over  the combined channel fading effect   of  atmospheric turbulence,  random fog, air-to-water interface, oceanic turbulence, and pointing errors.   We develop novel analytical expressions for the probability density function (PDF) and cumulative distribution function (CDF) of the resultant signal-to-noise ratio (SNR) in terms of bivariate Meijer-G and  Fox-H functions. We use the derived statistical results to analyze the system performance by providing  exact and asymptotic  results of the outage probability in terms of system parameters.  We use computer simulations to demonstrate the performance of  direct A2UW transmissions compared to the relay-assisted system.
\end{abstract}


\section{Introduction}\label{sec:introduction}
Optical wireless communication (OWC) is a potential technology for  underwater applications such as   oceanographic data collection, tactical surveillance, and  offshore explorations \cite{Zeng2017}. The OWC system performs exceedingly well for underwater communication achieving ultra-high data rate secured transmission at a  low-power consumption   compared to legacy technologies such as acoustic waves and radio-frequency (RF). 	Still, the performance of underwater OWC (UWOC) is limited by the oceanic turbulence caused by air bubbles levels and variation in the  temperature and pressure of the seawater in addition to other impairments such as scattering and absorption  \cite{Oubei2017,Jamali2018,Zedini2019}. 

There has been an increased research interest in studying the heterogeneous underwater-terrestrial network to   offload the underwater data using RF or OWC technologies. Cooperative relaying protocols such as amplify-and-forward (AF) and decode-and-forward (DF) are generally employed to interface the UWOC with terrestrial link using a buoy relay node \cite{Lei2020, Li2020, Li2021, Ansari_2021_TVT, Yang2021}.    In \cite{Li2021}, the authors analyzed the performance of an unmanned-aerial-vehicle (UAV) assisted  RF-UWOC system using both fixed-gain AF and DF relaying protocols.  The authors in \cite{Yang2021} used the OWC transmission over Gamma-Gamma turbulence  to study the fixed-gain AF relaying for the mixed UWOC-OWC communication system. 

In the aforementioned and related research, the deployment of  relaying devices, either floating buoys or mounted on a ship, may be challenging, and to the least, can impose security issues, especially if the underwater communication is carried out for tactical surveillance. Recently, there have been some preliminary  studies for direct transmissions, specifically towards the statistical modeling of  wavy water surface at the air-to-water interface\cite{Nabavi2019GC,Nabavi2019ICC,Lin2021}. The Birnbaum-Saunders distribution function was found to  be a better fit  to predict the statistical behavior of  the fading channel in the presence of random aquatic waves at the air-to-water interface \cite{Nabavi2019ICC}.   Further, Agheli \emph{et.al} \cite{Agheli2021JLT} presented a study for OWC transmission  directly to an underwater vehicle under the impact of air-to-water interface with log-normal turbulence model for underwater communication. However, they used  deterministic path loss and did not consider  atmospheric turbulence for the OWC link, which may overestimate the actual  performance. It should be mentioned that foggy weather conditions over the sea may randomly change the optical signal absorption, precluding deterministic path loss for the OWC link \cite{Esmail2017_Access,Rahman_Systems}. Moreover,  the log-normal  model for the underwater is limited for weak oceanic turbulence. To the best of the author's knowledge, generalized fading models for atmospheric and oceanic turbulence have not yet been studied for  the direct UWOC system. It requires novel approaches for statistically analyzing the system performance considering the different channel coefficients involved in the signal transmission.

In this paper, we investigate the feasibility of direct air-to-underwater (A2UW)  communication  from an over-the-sea OWC system  to an underwater submarine under generalized channel conditions. We analyze the statistical performance of the proposed scheme by considering the combined channel effect  of  atmospheric turbulence,  random fog, air-to-water interface, oceanic turbulence, and pointing errors. We consider the generalized Mal\'aga distribution  for the atmospheric turbulence, Gamma distributed attenuation coefficient for random fog, the BS distribution to model the aquatic waves at air-water interface, the mixture exponential–generalized gamma (EGG) distribution for the oceanic turbulence, and zero-boresight for random misalignment errors between transmitter and detector.   We develop novel analytical expressions for the probability density function (PDF) and cumulative distribution function (CDF) of the resultant signal-to-noise ratio (SNR) for the A2UW using the bivariate Fox's H-function.  We use the derived statistical results to analyze the system performance by providing  exact   outage probability  in terms of system parameters. We also analyze the outage performance asymptotically to derive diversity order  providing insights on the  system behavior  in the high SNR regime. We validate our derived analysis and use computer simulations to demonstrate the performance of the direct A2UW scheme with a comparison to the DF-based  relay-assisted system. The direct A2UW can provide acceptable performance for various scenarios of interest when the double fading effect  above-the-sea and  underwater channels is not severe.

\section{System and Channel Models}
We consider an  A2UW transmission scheme where an over-the-sea UAV directly communicates  with an underwater submarine. The UAV  transmits a source signal using the intensity modulation/direct-detection (IM/DD) technique for the destination located in  underwater.  The transmitted signal encounters  fading due to  atmospheric turbulence, random fog, oceanic turbulence, and pointing  errors. In addition to these channel fading impairments, the signal is affected by  erratic random and non-random aquatic waves at the air-water interface. We consider link without breaking waves (i.e., no bubbles) at the air-water interface. Thus, the electrical signal received (denoted by $y$) at the  underwater detector  can be expressed as
\begin{eqnarray}
y=h_{at}h_f h_{ws}H_wh_{ut}h_ps+n
\label{eq:rx1}
\end{eqnarray}
where $s$ is the transmitted signal, $h_{at}$ is the channel coefficient for the  atmospheric turbulence, $h_f$ models the randomness in the path gain due to the foggy condition, $ h_{ws}$ models the effect of scattering and reflection of the signal by water waves at the air-to-water interface, $H_{w}$  denotes the oceanic path gain, $h_{ut}$ is the oceanic turbulence, $h_p$  is the channel coefficient due to the pointing errors, and $n$ denotes the additive noise at the detector with variance $\sigma^2_n$. The parameterized values of the channel coefficients in \eqref{eq:rx1} are given in Table \ref{Simulation_Parameters}.

We consider the generalized Mal\'aga distribution  $\scriptscriptstyle f_{h_{at}}(x)= A_{\rm mg}\sum_{m=1}^{\beta_{{\scriptscriptstyle  M}}}a_{m} x K_{\alpha_M-m}(2\sqrt{\frac{\alpha_M\beta_M x}{g\beta_M+\Omega'}})$ to model the atmospheric turbulence, where the fading parameters  $\alpha_{{\scriptscriptstyle  M}}$, $\beta_{{\scriptscriptstyle  M}}$,  $A_{m\rm g}$, $a_m$, $g$, and $\Omega'$ are defined in \cite{malaga2011}.  The probability density function (PDF) of the foggy channel is  given as \cite{Esmail2017_Access}:
\begin{flalign}
f_{h_{f}}(x) = \frac{z^{k}}{\Gamma(k)}\left(\log\frac{1}{x}\right)^{k-1} x^{z-1},~0<x\leq 1
\label{eq:pdf_hf}
\end{flalign}
where $z=4.343/\beta_f d_{\rm air}$, $k>0$ is the shape parameter, $\beta_f>0$ is the scale parameter and $d_{\rm air}$ (in {\rm km}) is link distance from the water surface to the UAV. For the deterministic path gain, we can use the visibility model $h_f= e^{-\phi_{\rm air}d_{\rm air}}$, where $\phi_{\rm air}$ is the attenuation coefficient \cite{Kim2001}. The statistical characterization of the air-to-water interface is captured through  Birnbaum-Saunders (BS) distribution \cite{Nabavi2019ICC}:
\begin{eqnarray}
&f_{h_{ws}}(x) = \frac{1}{2\sqrt{2\pi}\alpha\beta}{\left[{\left(\frac{\beta}{x}\right)}^{1/2} + {\left(\frac{\beta}{x}\right)}^{3/2}\right]} \nonumber \\&
 \exp\left[-\frac{1}{2{\alpha}^2}\left(\frac{x}{\beta} + \frac{\beta}{x} - 2\right)\right]
\label{pdf_ws}
\end{eqnarray}
where $\alpha > 0$ and $\beta > 0$ are the shape and scale parameters to parameterize the BS distribution. The path gain for the oceanic turbulence can be approximated using Beer-Lambert law $H_w=e^{-\phi_{\rm water} d_{\rm water}}$ ($d_{\rm water}$ in \rm{m}) \cite{Agheli2021JLT}, where $\phi_{\rm water}$ is the extinction attenuation coefficient. Further, we use the recently proposed  mixture EGG distribution for the oceanic turbulence (caused by air bubbles and temperature gradient) \cite{Zedini2019}:
\begin{eqnarray}
f_{h_{ut}}(x) = \frac{\omega}{\lambda}\exp(-\frac{x}{\lambda})  + (1-\omega)\frac{cx^{ac-1}}{b^{ac}}\frac{\exp(-(\frac{x}{b})^c)}{\Gamma(a)}
\label{pdf_uw}
\end{eqnarray}
where $\omega$ is the mixture
coefficient of the distributions (i.e, $0<\omega>1$), $\lambda$ is the exponential distribution parameter, $a$, $b$, and $c$ are the generalized Gamma distribution parameters. Finally, we model the misalignment  between the transmit aperture of  UAV  and detector at the submarine  using the PDF $f_{h_{p}}(x) = \frac{\rho^2}{A_{0}^{\rho^2}}x^{\rho^{2}-1}$, where $A_0$ and $\rho$ are pointing error parameters \cite{Farid2007}.

\begin{figure}[t]
	\centering
	\vspace{-1.5cm}	
	\includegraphics[scale=0.6]{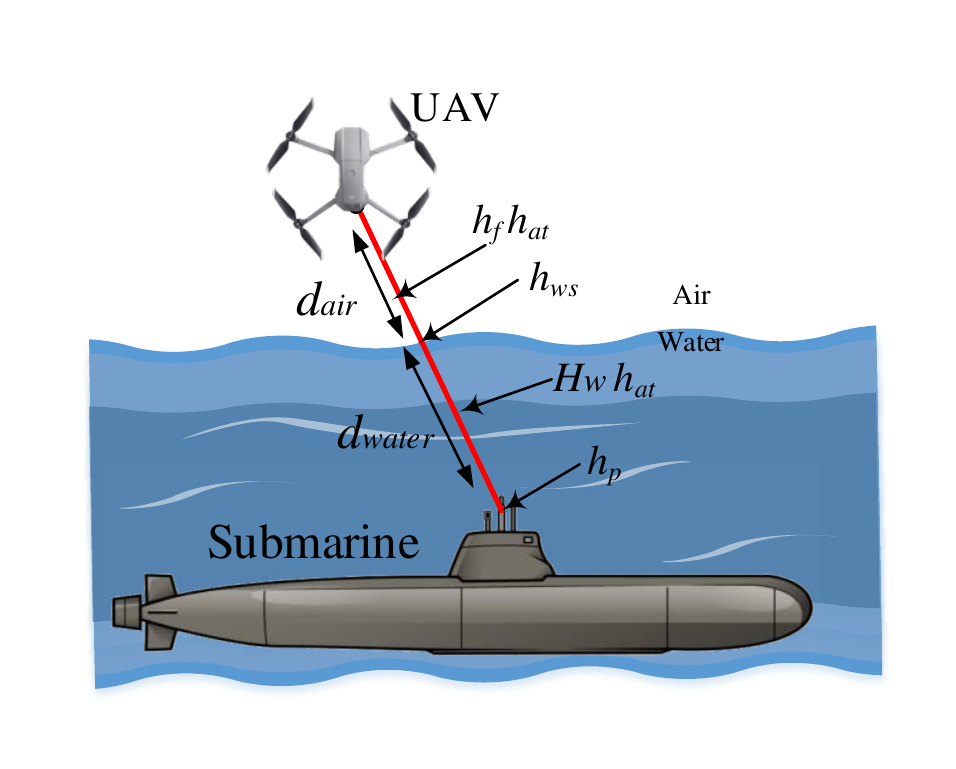}
		\vspace{-0.8cm}
		\caption{A schematic diagram for direct A2UW communication.}
	\label{model}
	
\end{figure}
\begin{figure*}
	\begin{eqnarray}
	& f_{\gamma}(\gamma)=\frac{z^k\rho^2A_{\rm{mg}}\omega}{8\sqrt{\pi}\gamma }\exp\left(\frac{1}{\alpha^2}\right) \sum_{m=1}^{\beta_m}b_m G_{1,0:1,0:4+k,1+k}^{0,1:0,1:0,4+k}\left(\begin{array}{c} \frac{1}{2}\\- \end{array}\left| \begin{array}{c} 1\\- \end{array}\right.\right.  \left. \left| \begin{array}{c} \zeta_1 \\\chi_1 \end{array} \right.  \left| 4\alpha^4,  \frac{2{\alpha}^2\beta(g\beta_m+\Omega')A_0\lambda}{\alpha_m\beta_m }\sqrt{\frac{\gamma_0}{\gamma}} \right. \right)\nonumber \\& + \frac{z^k\rho^2A_{\rm{mg}}\omega}{16\sqrt{\pi}\alpha^2\gamma}\exp\left(\frac{1}{\alpha^2}\right) \sum_{m=1}^{\beta_m}b_m G_{1,0:1,0:4+k,1+k}^{0,1:0,1:0,4+k}\left(\begin{array}{c} \frac{3}{2}\\- \end{array}\left| \begin{array}{c} 1\\- \end{array}\right. \right.  \left.\left| \begin{array}{c} \zeta_1 \\\chi_1 \end{array} \right.  \left|4\alpha^4,  \frac{2{\alpha}^2\beta(g\beta_m+\Omega')A_0\lambda}{\alpha_m\beta_m}\sqrt{\frac{\gamma_0}{\gamma}} \right. \right)  \nonumber \\&+ \frac{z^k\rho^2A_{\rm{mg}}}{8\sqrt{\pi}\gamma }\frac{1-\omega}{\Gamma(a)}\exp\left(\frac{1}{\alpha^2}\right)  \sum_{m=1}^{\beta_m}b_m H_{1,0:1,0;4+k,1+k}^{0,1:0,1;0,4+k} \left( \begin{array}{c} \zeta_2 \\ \chi_2 \end{array} \left| 4\alpha^4,  \frac{2{\alpha}^2\beta(g\beta_m+\Omega')A_0b}{\alpha_m\beta_m}\sqrt{\frac{\gamma_0}{\gamma}} \right. \right) 
	\nonumber \\&+\frac{z^k\rho^2A_{\rm{mg}}}{16\sqrt{\pi}\alpha^2\gamma }\frac{1-\omega}{\Gamma(a)}\exp\left(\frac{1}{\alpha^2}\right)  \sum_{m=1}^{\beta_m}b_m H_{1,0:1,0;4+k,1+k}^{0,1:0,1;0,4+k}\left( \begin{array}{c} \zeta_3\\ \chi_2 \end{array} \left| 4\alpha^4,  \frac{2{\alpha}^2\beta(g\beta_m+\Omega')A_0b}{\alpha_m\beta_m}\sqrt{\frac{\gamma_0}{\gamma}} \right. \right)
	\label{snr_pdf} 
	\end{eqnarray}
	where $\zeta_1=\{0, 1-\rho^2, 1-\alpha_m, 1-m, \{1-z\}_1^k\}$, $\chi_1=\{-\rho^2, \{-z\}_1^k\}$, $\zeta_2=\{(\frac{1}{2};1,1):(1,1);(1-a,\frac{1}{c}), (1-\rho^2,1), (1-\alpha_m,1), (1-m,1), (\{1-z\}_1^k,1) \}$, $\chi_2=\{-:-; (-\rho^2,1), (\{-z\}_1^k,1)\}$, and $\zeta_3=\{(\frac{3}{2};1,1):(1,1);(1-a,\frac{1}{c}), (1-\rho^2,1), (1-\alpha_m,1), (1-m,1), (\{1-z\}_1^k,1) \}$.
	\hrule
\end{figure*}
\begin{figure*}[t]
	\begin{eqnarray}
		&P_{\rm out}=\frac{z^k\rho^2A_{\rm{mg}}\omega}{4\sqrt{\pi} }\exp\left(\frac{1}{\alpha^2}\right) \sum_{m=1}^{\beta_m}b_m G_{1,0:1,0:5+k,2+k}^{0,1:0,1:1,4+k}\left(\begin{array}{c} \frac{1}{2}\\- \end{array}\left| \begin{array}{c} 1\\- \end{array}\right.\right. \left. \left| \begin{array}{c} \zeta_4\\\chi_4\end{array} \right. \right.   \left. \left| 4\alpha^4,  \frac{2{\alpha}^2\beta(g\beta_m+\Omega')A_0\lambda}{\alpha_m\beta_m }\sqrt{\frac{\gamma_0}{\gamma_{\rm th}}} \right. \right) \nonumber \\& +  \frac{z^k\rho^2A_{\rm{mg}}\omega}{8\sqrt{\pi}\alpha^2 }\exp\left(\frac{1}{\alpha^2}\right)  \sum_{m=1}^{\beta_m}b_m G_{1,0:1,0:5+k,2+k}^{0,1:0,1:1,4+k}\left(\begin{array}{c} \frac{3}{2}\\- \end{array}\left| \begin{array}{c} 1\\- \end{array}\right. \right.  \left.\left| \begin{array}{c} \zeta_4 \\\chi_4 \end{array} \right. \right. \left.\left|4\alpha^4, \frac{2{\alpha}^2\beta(g\beta_m+\Omega')A_0\lambda}{\alpha_m\beta_m}\sqrt{\frac{\gamma_0}{\gamma_{\rm th}}} \right. \right) \nonumber \\&+ \frac{z^k\rho^2A_{\rm{mg}}}{4\sqrt{\pi} }\frac{1-\omega}{\Gamma(a)}\exp\left(\frac{1}{\alpha^2}\right) \sum_{m=1}^{\beta_m}b_m H_{1,0:1,0;5+k,2+k}^{0,1:0,1;1,4+k}\left( \begin{array}{c} \zeta_5 \\ \chi_5 \end{array} \left| 4\alpha^4,  \frac{2{\alpha}^2\beta(g\beta_m+\Omega')A_0b}{\alpha_m\beta_m}\sqrt{\frac{\gamma_0}{\gamma_{\rm th}}} \right. \right) 
		\nonumber \\&+\frac{z^k\rho^2A_{\rm{mg}}}{8\sqrt{\pi}\alpha^2 }\frac{1-\omega}{\Gamma(a)}\exp\left(\frac{1}{\alpha^2}\right)  \sum_{m=1}^{\beta_m}b_m H_{1,0:1,0;5+k,2+k}^{0,1:0,1;1,4+k}\left( \begin{array}{c} \zeta_6\\ \chi_5 \end{array} \left| 4\alpha^4,  \frac{2{\alpha}^2\beta(g\beta_m+\Omega')A_0b}{\alpha_m\beta_m}\sqrt{\frac{\gamma_0}{\gamma_{\rm th}}} \right. \right)
		\label{snr_cdf} 
	\end{eqnarray}
	where $\zeta_4=\{0, 1-\rho^2, 1-\alpha_m, 1-m, \{1-z\}_1^k, 1\}$, $\chi_4=\{0, -\rho^2, \{-z\}_1^k\}$, $\zeta_5=\{(\frac{1}{2};1,1):(1,1);(1-a,\frac{1}{c}), (1-\rho^2,1), (1-\alpha_m,1), (1-m,1), (\{1-z\}_1^k,1),(1,1) \}$, $\chi_5=\{-: (0,1); (-\rho^2,1), (\{-z\}_1^k,1)\}$, and $\zeta_6=\{(\frac{3}{2};1,1):(1,1);(1-a,\frac{1}{c}), (1-\rho^2,1), (1-\alpha_m,1), (1-m,1), (\{1-z\}_1^k,1),(1,1) \}$.
	\hrule
\end{figure*}

\begin{figure*}
	\begin{eqnarray}
	&P_{\rm out}^{\infty}=\frac{z^k\rho^2A_{\rm{mg}}\omega}{4\sqrt{\pi} }\exp\left(\frac{1}{\alpha^2}\right) \sum_{m=1}^{\beta_m}b_m \sum_{i=1}^{4+k}\Gamma\left(\frac{3}{2}+\mathcal{A}_i\right)\frac{\prod_{j=1,j\neq i}^{4+k}\Gamma(\mathcal{A}_i-\mathcal{A}_j)\Gamma(1-\mathcal{A}_i)}{\prod_{2}^{2+k}\Gamma(\mathcal{A}_i+\mathcal{B}_j)\Gamma(2-\mathcal{A}_j)}\left(\frac{2{\alpha}^2\beta(g\beta_m+\Omega')A_0\lambda}{\alpha_m\beta_m}\sqrt{\frac{\gamma_0}{\gamma}}\right)^{\mathcal{A}_i-1}\nonumber\\&+\frac{z^k\rho^2A_{\rm{mg}}\omega}{8\sqrt{\pi}\alpha^2 }\exp\left(\frac{1}{\alpha^2}\right)  \sum_{m=1}^{\beta_m}b_m\sum_{i=1}^{4+k}\Gamma\left(\frac{1}{2}+\mathcal{A}_i\right)\frac{\prod_{j=1,j\neq i}^{4+k}\Gamma(\mathcal{A}_i-\mathcal{A}_j)\Gamma(1-\mathcal{A}_i)}{\prod_{2}^{2+k}\Gamma(\mathcal{A}_i+\mathcal{B}_j)\Gamma(2-\mathcal{A}_j)}\left(\frac{2{\alpha}^2\beta(g\beta_m+\Omega')A_0\lambda}{\alpha_m\beta_m}\sqrt{\frac{\gamma_0}{\gamma}}\right)^{\mathcal{A}_i-1}\nonumber\\&+\frac{z^k\rho^2A_{\rm{mg}}}{4\sqrt{\pi} }\frac{1-\omega}{\Gamma(a)}\exp\left(\frac{1}{\alpha^2}\right) \sum_{m=1}^{\beta_m}b_m \sum_{i=1}^{4+k}\sum_{i=1}^{4+k}\frac{1}{\mathcal{Q}_i}\Gamma\left(\frac{3}{2}+\mathcal{P}_i\right)\frac{\Gamma\left(-(\mathcal{P}_i-1)\frac{\mathcal{T}_j}{\mathcal{Q}_i}\right)\prod_{j=1,j\neq i}^{4+k}\Gamma\left(1+\mathcal{P}_j+(\mathcal{P}_i-1)\frac{\mathcal{Q}_j}{\mathcal{Q}_i}\right)}{\Gamma\left(1-(\mathcal{P}_i-1)\frac{1}{\mathcal{Q}_i}\right)\prod_{2}^{2+k}\Gamma\left(1-\mathcal{S}_j+(\mathcal{P}_i-1)\frac{\mathcal{T}_j}{\mathcal{Q}_i}\right)}\nonumber\\&\left(\frac{2{\alpha}^2\beta(g\beta_m+\Omega')A_0 b}{\alpha_m\beta_m}\sqrt{\frac{\gamma_0}{\gamma}}\right)^{(\mathcal{P}_i-1)/\mathcal{Q}_i}+\frac{z^k\rho^2A_{\rm{mg}}}{8\sqrt{\pi} \alpha^2}\frac{1-\omega}{\Gamma(a)}\exp\left(\frac{1}{\alpha^2}\right) \sum_{m=1}^{\beta_m}b_m \sum_{i=1}^{4+k}\sum_{i=1}^{4+k}\frac{1}{\mathcal{Q}_i}\nonumber\\&\Gamma\left(\frac{1}{2}+a_i\right)\frac{\Gamma\left(-(a_i-1)\frac{\beta_j}{\mathcal{P}_i}\right)\prod_{j=1,j\neq i}^{4+k}\Gamma\left(1+\mathcal{P}_j+(\mathcal{P}_i-1)\frac{\mathcal{Q}_j}{\mathcal{Q}_i}\right)}{\Gamma\left(1-(\mathcal{P}_i-1)\frac{1}{\mathcal{Q}_i}\right)\prod_{2}^{2+k}\Gamma\left(1-\mathcal{S}_j+(\mathcal{P}_i-1)\frac{\mathcal{T}_j}{\mathcal{Q}_i}\right)}\left(\frac{2{\alpha}^2\beta(g\beta_m+\Omega')A_0 b}{\alpha_m\beta_m}\sqrt{\frac{\gamma_0}{\gamma}}\right)^{(\mathcal{P}_i-1)/\mathcal{Q}_i}
	\label{snr_cdf_asymp} 
	\end{eqnarray}
		where $\mathcal{A}_i=\mathcal{A}_j= \{0, 1-\rho^2, 1-\alpha_m, 1-m, \{1-z\}_1^k, 1\}$, $\mathcal{B}_i=\mathcal{B}_j=\{0, -\rho^2, \{-z\}_1^k\}$, $\mathcal{P}_i=\mathcal{P}_j=\{1-a,1-\rho^2, 1-\alpha_m, 1-m, \{1-z\}_1^k,1\}$, and $\mathcal{Q}_i=\mathcal{Q}_j=\{\frac{1}{c}, 1, 1, 1, \{1\}_1^k, 1\},\mathcal{S}_i=\mathcal{S}_j=\{-\rho^2, \{-z\}_1^k\},\mathcal{T}_i=\mathcal{T}_j=\{1, \{1\}_1^k\}$.
	\hrule
\end{figure*}
\section{Statistical Characterization}
In this section, we statistically characterize the direct transmission by deriving PDF and CDF of the SNR  for the combined channel effect consisting of  over-the-sea and underwater fading channels.  

We define $\gamma=\gamma_0 |h|^2$ as the SNR for the system model of \eqref{eq:rx1}, where the combined channel $h = h_{at}h_f h_{ws}h_{ut}h_p$ with $\gamma_0=2P^2_sR^2 H^2_w/\sigma^2_{n} $,  $P_s$ is the average optical transmitted power, and $R$ is the responsitivity. In the following theorem, we develop PDF of the SNR for the direct transmission scheme as given in  \eqref{eq:rx1}, considering fading models of atmospheric turbulence,  random fog, air-to-water interface, oceanic turbulence, and pointing errors. It should be mentioned that direct  application of the product of random variables is not readily applicable to derive the PDF of SNR for the A2UW.
\begin{my_theorem}
If  $\{\alpha_{{\scriptscriptstyle  M}}$, $\beta_{{\scriptscriptstyle  M}}$,  $A_{m\rm g}$, $a_m$, $g$, $\Omega'\}$ models the atmospheric turbulence,  $\{z=4.343/\beta_f d$, $k\}$ models the random fog, $\{\alpha, \beta\}$ models the air-to-water interface,   $\{\omega, \lambda, a, b, c\}$ models the oceanic turbulence, and $\{\rho, A_0\}$ models the pointing errors, then the PDF of SNR $\gamma$ for the direct A2UW scheme is given in 	\eqref{snr_pdf} (see top of the next page).
\end{my_theorem}

\begin{IEEEproof}
	The proof is presented in Appendix A.
	\end{IEEEproof}
It should be mentioned that  standard functions are available in MATLAB and MATHEMATICA   to compute  bivariate Meijer-G and Fox-H functions.

\section{Outage Probability}
We use the derived statistical results to study the outage probability of the system as a performance metric to compare with the relay-assisted transmissions. Note that other statistical performance metrics such as average BER and ergodic capacity can be similarly derived.   

\subsection{Exact Analysis}
We can use \eqref{snr_pdf} to develop an exact outage probability of the considered system  for a given threshold SNR $\gamma_{\rm th}$ as $P_{\rm out}= P[\gamma < \gamma_{\rm th }]$. Thus, we substitute \eqref{snr_pdf}  in   $P_{\rm out}=\int_{0}^{\gamma}f_{\gamma}(\gamma)d\gamma$ and apply standard Mathematical as used in deriving the PDF    to get the outage probability in \eqref{snr_cdf}.

\subsection{Asymptotic  Analysis}
Although \eqref{snr_cdf} can provide outage probability over a wide range of $\gamma_0$,  it is desirable to analyze the outage probability asymptotically at a high SNR  $\gamma_0\to \infty$. Thus, we use the asymptotic result of a single-variate Fox-H function in  \cite[Th. 1.11]{Kilbas_FoxH} and multivariate Fox-H \cite{Rahama2018} to express the outage probability in the high SNR regime as presented in 	\eqref{snr_cdf_asymp}.

\subsection{Diversity Order}
Compiling  the dominant terms of $\gamma_0$ in \eqref{snr_cdf_asymp} and using the definitions $\mathcal{A}_i$,  $\mathcal{P}_i$, and  $\mathcal{Q}_i$, the diversity order for the A2UW can be expressed as $G^{}_{\rm out} = \min{\{\frac{1}{2}, \frac{\alpha_m}{2}, \frac{\beta_m}{2}, \frac{z}{2}, \frac{ac}{2}, \frac{\rho^2}{2}\}}$. It can be seen that the diversity order  in \cite{Zedini2019} \cite{Yang2021} becomes a special case of our generalized model, which includes random fog and air-to-water interface in addition to the atmospheric and oceanic turbulence with pointing errors. The  diversity order reveals several interesting behaviors of the outage probability for the direct A2UW transmission, specifically: (i) it is independent of the fading parameters of the BS distribution modeled for the air-to-water interface; and (ii) the diversity order is  $\frac{1}{2}$ or  $\frac{z}{2}$ if $z(=\frac{4.343}{\beta_f d_{\rm air}}) <1$ depending on the density fog and $d_{\rm air}$ since measurement data reveals that typically $ac>1$, $\alpha_M>1$, $\beta_M>1$, and $\rho>1$ can be maintained with higher beam-width to reduce the impact of pointing errors. 

\begin{table}[t]	
	\renewcommand{\arraystretch}{01}
	\caption{Simulation Parameters}
	\centering
	\begin{tabular}{| p{4cm}| p{3.8cm}|}
		\hline 	
		
		Noise &$\sigma_n^2=10^{-14}~\rm {A^2/GHz}$ \\ \hline	
		Light Fog  &  $k=2.32$, $\beta_f=13.12$ \\ \hline
		Moderate Fog  &  $k=5.49$, $\beta_f= 12.06$ \\ \hline
		Atten. Coeff. (Haze)   &  $\phi_{\rm air}=0.98$ \cite{Kim2001} \\ \hline
		Atten. Coeff. (UW)	& $\phi_{\rm water}=21.79 ~{\rm dB/km}$ \cite{Agheli2021JLT}
		\\ \hline
		EGG-1  (Weaker)	& $\omega=0.21$, $\lambda=0.329$  $a=1.429$, $b=1.181$, $c=17.198$\\ \hline
		EGG-2 (Strong) 	& $\omega=0.458$, $\lambda=0.344$  $a=1.042$, $b=1.576$, $c=35.942$\\ \hline
		BS Distribution  & $\alpha=\{0.3, 07\}$, $\beta=\{1,1.5\}$
		\\ \hline
		
	\end{tabular}
	\label{Simulation_Parameters}	 
\end{table}

\begin{figure*}[t]
	\subfigure[PDF of combined  air-to-water interface and oceanic turbulence..]
{\includegraphics[scale=0.20]{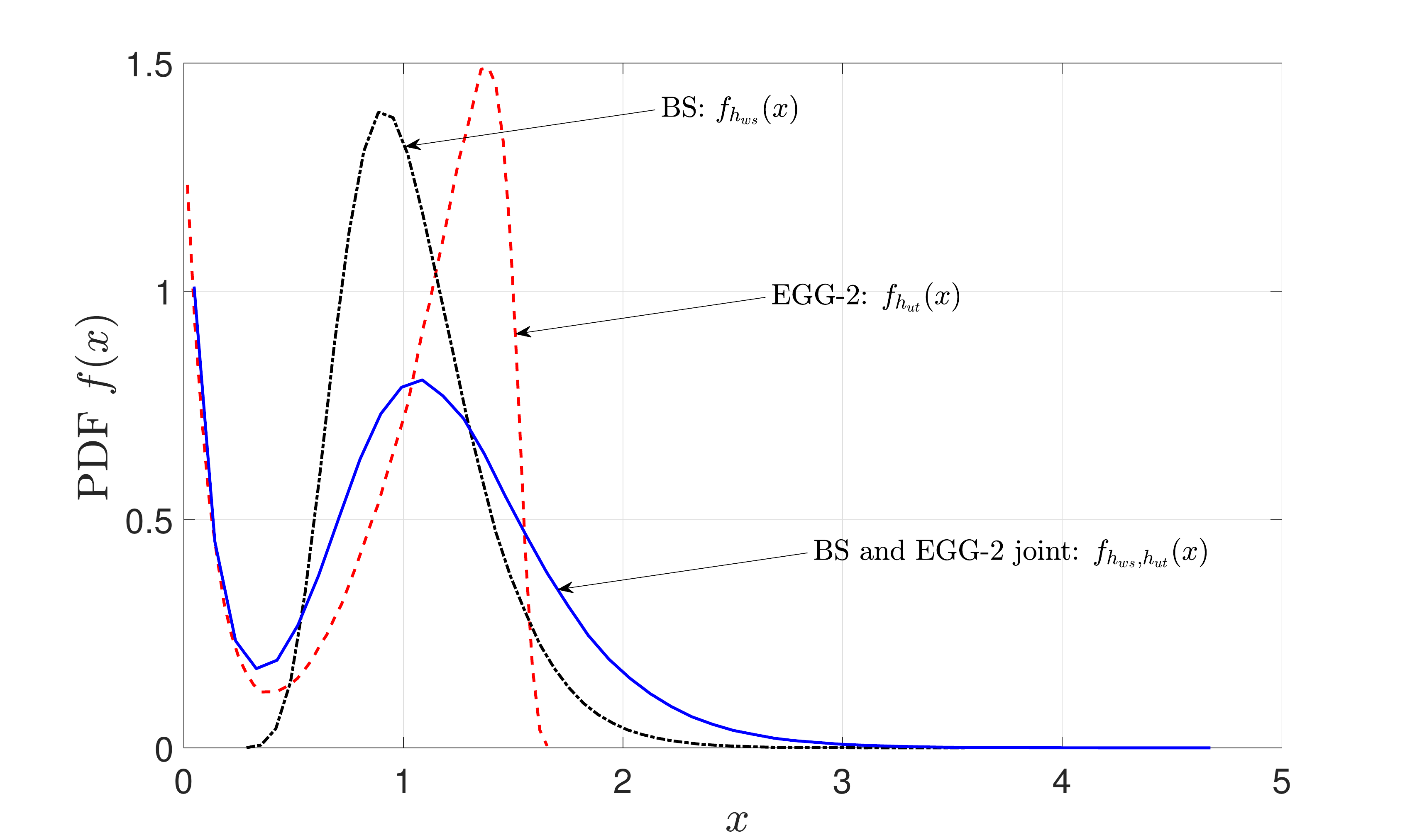}}
	\subfigure[EGG-2, light fog, weak atmospheric turbulence, $\rho=5$, $A_0=0.0032$.]
	{\includegraphics[scale=0.22]{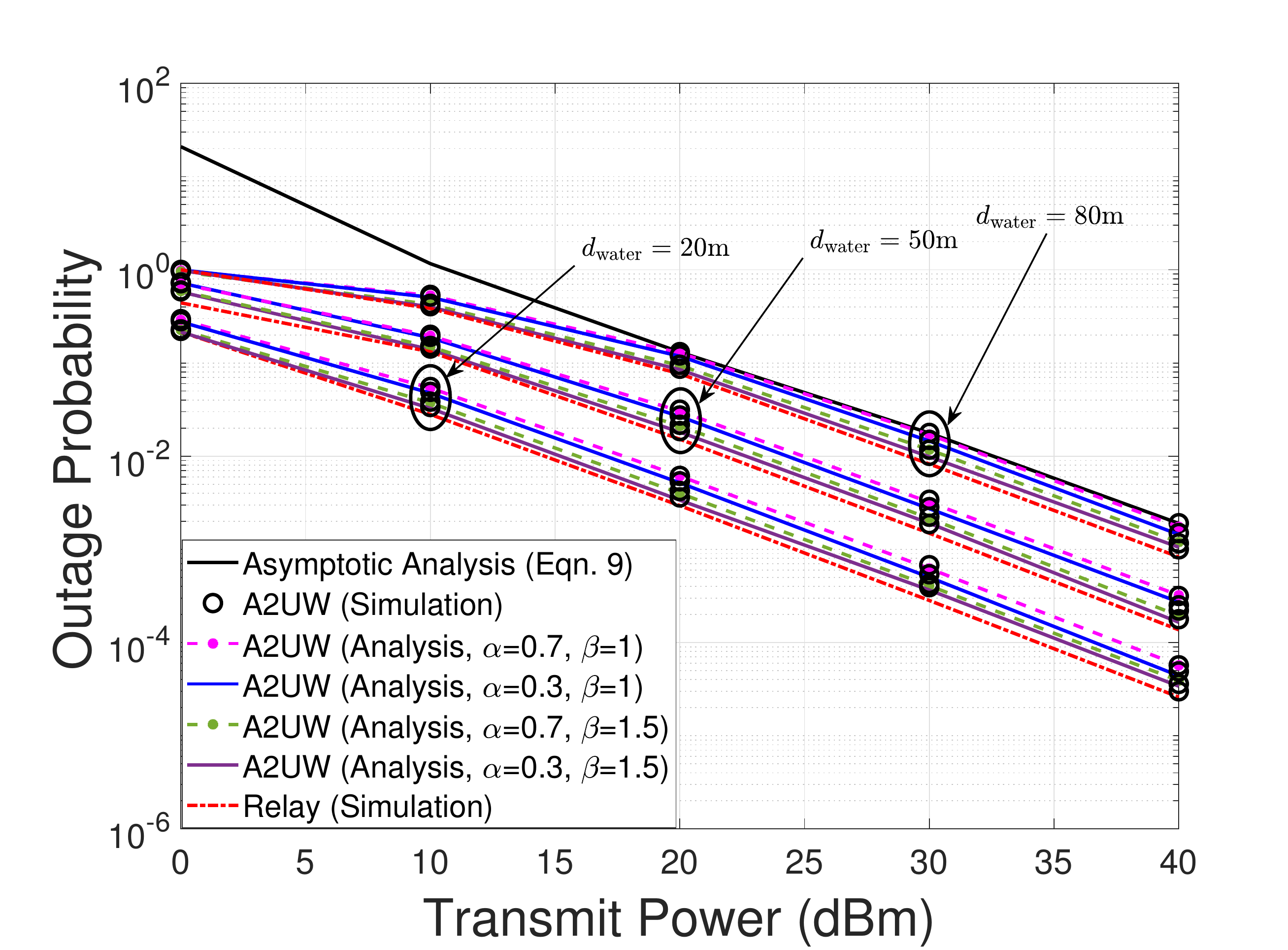}}
	\subfigure[ $\alpha=0.3$, $\beta=1$; $d_{\rm air}= 20$\mbox{m}, $d_{\rm water}= 50$\mbox{m}.]
	{\includegraphics[scale=0.22]{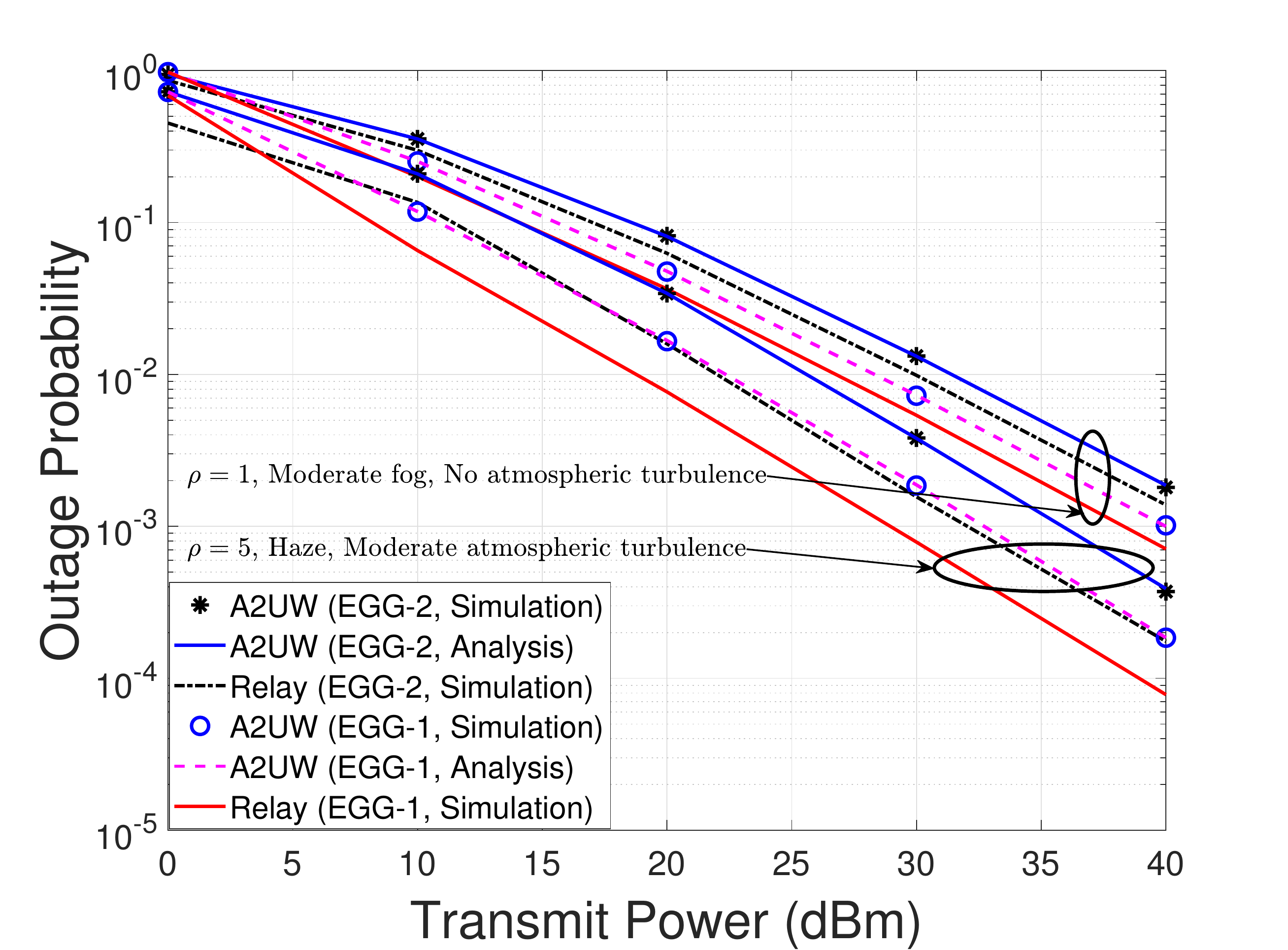}} 
	\caption{Direct A2UW transmission with a comparison to the relay-assisted system. }
	\label{out_prob}
\end{figure*}

\section{Simulation Results and Discussions}
We use MATLAB software to validate our analysis and demonstrate the performance of the direct A2UW transmission considering various system and channel configurations. We also compare the performance with the DF-based relaying  to better assess use cases for the proposed scheme. We fix the air-to-water surface link distance $d_{\rm air}$ from $20$\mbox{m} (for example, a UAV hovering over the sea), and consider  underwater link range  $d_{\rm water}$ from $20$\mbox{m} to $80$\mbox{m}. We use three atmospheric turbulence (weak, medium, and strong)  \cite{malaga2011} and consider low $\rho=5$ and high $\rho=1$ pointing errors with $A_0=0.0032$. Other  simulation parameters are listed in Table \ref{Simulation_Parameters}.

First, we demonstrate the effect of fading due to the aquatic waves at the air-to-water interface by comparing PDF of the combined BS and EGG distributions with the EGG, as shown in  Fig.~\ref{out_prob}(a). The figure depicts that the BS distribution  deteriorates  random channel conditions for the direct A2UW transmission since the probability of getting the peak value decreases as compared with the EGG. However, the PDF plot of $f_{h_{ws}h_{ut}}(x)$ reveals that the average value of the combined channel may increase due to higher non-zero values compared with the EGG alone.

Next, we compare the outage probability of the direct A2UW transmission with the DF relay system  at three different underwater distances $d_{\rm water}=20$\mbox{m}, $d_{\rm water}=50$\mbox{m}, and $d_{\rm water}=80$\mbox{m}, as shown in Fig.~\ref{out_prob}(b).  The atmospheric turbulence is considered to be weak, which is a reasonable assumption at  such a short distance.  The figure shows that the direct A2UW transmission  achieves performance close to the relay-assisted system  for different underwater link distances. However, there is a gap in the performance $2$\mbox{dBm} when the BS parameters become $\alpha=0.7$ and $\beta=1$.  Fig.~\eqref{out_prob} also demonstrates that the effect of BS parameters ($\alpha$, $\beta$) on the outage probability is marginal, and the slope remains constant, which confirms our analysis for the diversity order. 

Finally, in Fig.~\ref{out_prob}(c), we demonstrate the interplay of oceanic turbulence, pointing errors, atmospheric turbulence, and  fog density on the outage performance. The figure shows that  the A2UW performs close to the relayed system for both strong and weaker oceanic turbulence scenarios without atmospheric turbulence. However, the direct transmission incurs a penalty of $4$\mbox{dBm} transmit power to achieve the exact outage probability of the relayed system in the presence of  atmospheric turbulence. The degradation in the A2UW performance  occurs due to the double fading effect of turbulence and   path gain of two mediums for the direct A2UW transmissions  compared with the dominant single link performance using the DF protocol. The figure also confirms the relative  effect of pointing errors ($\rho=1$ and $\rho=5$) and intensity of oceanic turbulence (EGG-1 and EGG-2)   on  wireless transmissions.

We can conclude  that the direct A2UW achieves acceptable performance for various scenarios of interest and performs  close to the relay-assisted transmissions when the double fading effect caused by the cascading  of  above-the-sea and  underwater channels is not severe. Nevertheless, relay-assisted transmission is optimal in the mixed OWC-UWOC transmission. Experimental demonstration are further required to support the theoretical analysis developed in this paper.  We envision that the proposed direct A2UW transmission may further substantiate  underwater  communications using optical carriers.

\section* {Appendix A: PDF of SNR }
To derive the PDF of the product of $5$ random variables, we express $h = h_1 h_f h_{ws}h_{ut}$, where $h_1= h_{at}h_p$. The PDF of $h_1$ is given in \cite{Ansari2016}
	\begin{align}
		\begin{split}
			f_{h_{1}}(x)= \frac{\rho^{2}A_{\rm mg}}{2 x}\sum_{m=1}^{\beta_{{\scriptscriptstyle  M}}}b_{m}G_{1,3}^{3,0}\left(\frac{\alpha_{{\scriptscriptstyle  M}}\beta_{{\scriptscriptstyle  M}}}{g\beta_{{\scriptscriptstyle  M}}+\Omega^{'}}\frac{x}{A_{0}} \left|
			\begin{array}{c}
				\rho^{2}+1 \\
				\rho^{2},\alpha_{{\scriptscriptstyle  M}},m\\
			\end{array}
			\right.\right)
			\label{eq:PDF_Malaga_pe2}
		\end{split}
	\end{align}	
	Next, we express $h=h_2h_{ws}h_{ut}$, where $h_2= h_1 h_f$. Substituting  \eqref{eq:pdf_hf} and \eqref{eq:PDF_Malaga_pe2}  in	$f_{h_2}(x)=\int_{x}^{\infty}\frac{1}{|h_{1}|}f_{h_{f}}({h_2|h_{1}}) f_{h_{1}}(h_{1})dh_{1}$, we get
	\begin{eqnarray}
		&f_{h_2}(x) = \frac{\rho^{2}A_{\rm mg}}{2 h_1}x^{z-1}\int_{x}^{\infty}h_{1}^{-z-1}\left[\ln \left(\frac{h_{1}}{x}\right)\right]^{k-1} \nonumber \\&\sum_{m=1}^{\beta_{{\scriptscriptstyle  M}}}b_{m}G_{1,3}^{3,0}\left(\frac{\alpha_{{\scriptscriptstyle  M}}\beta_{{\scriptscriptstyle  M}}}{g\beta_{{\scriptscriptstyle  M}}+\Omega^{'}}\frac{h_1}{A_{0}} \left|
		\begin{array}{c}
			\rho^{2}+1 \\
			\rho^{2},\alpha_{{\scriptscriptstyle  M}},m\\
		\end{array}
		\right.\right) dh_{1}
		\label{eq_int}
	\end{eqnarray}
	
	Using the definition of Meijer-G function and substituting $\ln(h_{1}/x)=t$, we solve  the  inner integral $\int_{0}^{\infty}t^{k-1}\exp(-(z-s)t)dt=\frac{\Gamma(k)}{(z-s)^{k}}=\Gamma(k)\left[\frac{\Gamma\left(z-s\right)}{\Gamma\left(1+z-s\right)}\right]^{k}$ in terms of the Gamma function, and apply the definition of the Meijer-G function to  get the PDF of $h_2$:
	\begin{eqnarray}
		&f_{h_{2}}(x)=\frac{z^k\rho^2A_{\rm{mg}}}{2 x}  \sum_{m=1}^{\beta_m}b_m\nonumber \\&G_{1+k,3+k}^{3+k,0} \left(\begin{array}{c} \rho^2+1, \{z+1\}_1^k\\ \rho^2, \alpha_m, m, \{z\}_1^k\end{array} \left| \frac{\alpha_m\beta_m x}{(g\beta_m+\Omega')A_0}\right.\right)
		\label{eq:pdf_fpt}
	\end{eqnarray}	
	Note that the PDF in \eqref{eq:pdf_fpt} can be verified using the unified expression in \cite{chapala2021}.
	
Next, we express $h=h_3h_{ut}$, where $h_3= h_2 h_{ws}$. The product distribution of $h_3$ can be expressed as
	\begin{eqnarray}
		f_{h_3}(x)=\int_{0}^{\infty}\frac{1}{|h_{ws}|}f_{h_{2}}(h_3/h_{ws})f_{h_{ws}}(h_{ws})dh_{ws}
		\label{combine_pdf_fpt_ws}
	\end{eqnarray}
	Since the direct use  of \eqref{pdf_ws} in \eqref{combine_pdf_fpt_ws} becomes intractable, we convert the PDF of BS distribution in \eqref{pdf_ws}  as
	\begin{eqnarray}
		& f_{h_{ws}}(h_{ws}) = \frac{1}{2\sqrt{2\pi}\alpha\beta} \exp\left(\frac{1}{\alpha^2}\right){\left(\frac{\beta}{h_{ws}}\right)}^{1/2}  \nonumber \\&   G_{0,1}^{1,0}\left(\begin{array}{c} -\\0 \end{array}\middle \vert \frac{h_{ws}}{2{\alpha}^2\beta}\right)
		G_{0,1}^{1,0}\left(\begin{array}{c} -\\0 \end{array}\middle \vert \frac{\beta}{2{\alpha}^2h_{ws}}\right) \nonumber \\& + \frac{1}{2\sqrt{2\pi}\alpha\beta}\exp\left(\frac{1}{\alpha^2}\right){\left(\frac{\beta}{h_{ws}}\right)}^{3/2}  \nonumber \\&  G_{0,1}^{1,0}\left(\begin{array}{c} -\\0 \end{array}\middle \vert
		\frac{h_{ws}}{2{\alpha}^2\beta}\right) G_{0,1}^{1,0}\left(\begin{array}{c} -\\0 \end{array}\middle \vert \frac{\beta}{2{\alpha}^2h_{ws}}\right)
		\label{eq:ws_meijer}
	\end{eqnarray} 
	Substituting \eqref{eq:pdf_fpt} and \eqref{eq:ws_meijer} in \eqref{combine_pdf_fpt_ws} with the identity $G_{p,q}^{m,n}\left(\begin{array}{c} a_p \\b_q \end{array} \left| z\right. \right) = G_{q,p}^{n,m}\left(\begin{array}{c} 1-b_q \\1-a_p \end{array} \left| z^{-1}\right. \right)$  and  applying the identity \cite[eq. $07.34.21.0081.01$]{Wolfram}, we get
	\begin{eqnarray}
		& f_{h_3}(x) = \frac{z^k\rho^2A_{\rm{mg}}}{4\sqrt{\pi} x}\exp\left(\frac{1}{\alpha^2}\right)  \sum_{m=1}^{\beta_m}b_m G_{1,0:1,0:3+k,1+k}^{0,1:0,1:0,3+k}\nonumber \\&\left(\begin{array}{c} \frac{1}{2}\\- \end{array}\left| \begin{array}{c} 1\\- \end{array}\right.\right. \left. \left| \begin{array}{c} \zeta_7 \\\chi_7 \end{array} \right.  \left| 4\alpha^4,  \frac{2{\alpha}^2\beta(g\beta_m+\Omega')A_0}{\alpha_m\beta_m x} \right. \right) \nonumber \\& +  \frac{z^k\rho^2A_{\rm{mg}}}{8\sqrt{\pi}\alpha^2 x}\exp\left(\frac{1}{\alpha^2}\right) \sum_{m=1}^{\beta_m}b_m  G_{1,0:1,0:3+k,1+k}^{0,1:0,1:0,3+k}\nonumber \\&\left(\begin{array}{c} \frac{3}{2}\\- \end{array}\left| \begin{array}{c} 1\\- \end{array}\right. \right. \left.\left| \begin{array}{c} \zeta_7 \\\chi_7 \end{array} \right.  \left|4\alpha^4,  \frac{2{\alpha}^2\beta(g\beta_m+\Omega')A_0}{\alpha_m\beta_m x} \right. \right)
		\label{pdf_combine_fpt_ws}
	\end{eqnarray}
	where $\zeta_7=\{1-\rho^2, 1-\alpha_m, 1-m, \{1-z\}_1^k\}$, $\chi_7=-\rho^2, \{-z\}_1^k$.	Finally, we use  \eqref{pdf_uw} and \eqref{pdf_combine_fpt_ws} in $ 
		f_{h}(x)=\int_{0}^{\infty}\frac{1}{|h_{ut}|}f_{h_3}(h/h_{ut})f_{h_{ut}}(h_{ut})dh_{ut}$, and apply the line integral definition of Meijer-G and fox-H functions to get the PDF of  $h=h_3 h_{ut}$ by solving  four inner integrals as $I_1=I_2=\int_{0}^{\infty}h_{ut}^{s_2}\exp\left(-\frac{h_{ut}}{\lambda}\right)dh_{ut}=\lambda^{1+s_2}\Gamma(1+s_2)$, and $I_3=I_4=\int_{0}^{\infty}h_{ut}^{ac+s_2-1}\exp\left(-\left(\frac{h_{ut}}{\lambda}\right)^{c}\right)dh_{ut}=\frac{\Gamma\left(a+\frac{s_2}{c}\right)}{c\lambda^{ac+s_2}}$, and apply the definitions of bivariate Meijer-G and Fox-H functions \cite{Mittal_1972} with the transformation $h=\sqrt{\frac{\gamma}{\gamma_0}}$ to get the PDF of SNR $\gamma$ in \eqref{snr_pdf}.

	\bibliographystyle{IEEEtran}
	\bibliography{wcl}
\end{document}